\documentclass[twocolumn]{revtex4}
\usepackage[dvipdfm]{graphicx}
\usepackage{amsmath,amssymb}
\usepackage{bm}
\usepackage{booktabs}
\usepackage{comment}
\begin{document}

\title{Numerical Nahm transform for 2-caloron solutions}
\author{
Daichi Muranaka$^1$, Atsushi Nakamula$^2$, Nobuyuki Sawado$^1$ and Kouichi Toda$^3~^4$\\
~\\
\emph{
$\!\,^1$Department of Physics, Tokyo University of Science, 2641 Yamazaki, Noda, Chiba, 278-8510, Japan\\
$\!\,^2$Department of Physics, School of Science, Kitasato University, Sagamihara, 228-8555, Japan\\
$\!\,^3$Department of Mathematical Physics, Toyama Prefectural University,
Kurokawa 5180, Imizu, Toyama, 939-0398, Japan\\
$\!\,^4$Research and Education Center for Natural Sciences, 
Hiyoshi Campus, Keio University, 
4-1-1 Hiyoshi, Kouhoku-ku, Yokohama, 223-8521, 
Japan
}
}
\begin{abstract}
A new numerical method for performing the Nahm transform for charge $k=2$ caloron is presented.
The Weyl equations with boundary impurities are solved directly and the determination of the appropriate
basis to the linear system is established.
The action densities of the 2-calorons with 10 moduli parameters are shown.
\end{abstract}
\maketitle

\section{Introduction}

%%%%%%%%%%%%%%%%%%%%%%%%%%%%%%%%%%%%%%%%%%%%%%%%%%%%%%%%%%%%%%%

Calorons are finite action (anti-)self-dual ((A)SD) solutions of the Yang-Mills gauge theory on $\mathbb{R}^3\times S^1$.
Along with their periodic instanton picture, 
they can be interpreted as the compound objects of constituent monopoles from the perspective of 
loop group gauge theories \cite{Garland:1988bv, Norbury:1998ay}. 
The intriguing feature of calorons is that they can be endowed with  a non-trivial holonomy around the periodic direction,
 which will serve as an approximation to Skyrmions \cite{Atiyah:1989dq}.
Hence, calorons give a connection between instantons, monopoles and Skyrmions. 

%%%%%%%%%%%%%%%%%%%%%%%%%%%%%%%%%%%%%%%%%%%%%%%%%%%%%%%%%%%%%%%

There has been formulated a systematic method to construct instantons, 
the Atiyah-Drinfeld-Hitchin-Manin (ADHM) construction \cite{Atiyah:1978ri}.
Nahm  has applied the ADHM construction for calorons as well as monopoles \cite{Nahm:1983sv},
known as the Nahm construction.
In this formulation, gauge fields of calorons can be obtained by solving a one-dimensional 
Weyl equation on finite intervals with ``impurities" at the boundaries.
The dual gauge connection in Weyl operators and the impurities are called the bulk Nahm data and the 
boundary Nahm data, respectively.
The transformation from the Nahm data to the corresponding gauge connections is called the Nahm transform.
Since the Nahm data are defined in terms of quite complicated functions, 
we need numerical analysis to perform the Nahm transform, in general.

%%%%%%%%%%%%%%%%%%%%%%%%%%%%%%%%%%%%%%%%%%%%%%%%%%%%%%%%%%%%%%%

The Nahm transform for SU(2) caloron of instanton charge $k=1$ with non-trivial holonomy 
was, however, studied analytically \cite{Kraan:1998pm,Lee:1998bb}.
For the higher charges, some exact calorons of $k=2$ were found in \cite{Bruckmann:2004nu}. 
The authors demonstrated that the Nahm data written by the standard Jacobi 
elliptic functions works well for the case of calorons.
The numerical Nahm transform has been discussed
in somewhat different context, \textit{e.g.}, \cite{GonzalezArroyo:1998cq,GarciaPerez:1999bc}.

%%%%%%%%%%%%%%%%%%%%%%%%%%%%%%%%%%%%%%%%%%%%%%%%%%%%%%%%%%%%%%%

In this Letter, we perform the numerical Nahm transform for 
SU(2) calorons of instanton charge $k=2$, referred to as 2-calorons, in detail.
For a given instanton charge, calorons are classified by the constituent monopole charge and mass, respectively.
Here we consider the case of monopole charge $(m_1, m_2)=(2,2)$ and mass $(2\mu, \mu_0-2\mu)=(0,\mu_0)$,
which corresponds to the 2-caloron without net magnetic charge, and accompanied with trivial holonomy.
Recently, the 2-caloron Nahm data of monopole charge $(m_1, m_2)=(2,2)$ of arbitrary mass
 with  16 moduli parameters has been proposed in \cite{Nakamula:2009sq}.
Since the dimension of framed moduli space of SU(2) $k$-caloron is $8k$ \cite{Etesi-Jardim},  this 2-caloron Nahm data gives the most 
general gauge fields.
We carry out here the Nahm transform for the 10 parameter subset of the 2-caloron Nahm data whose holonomy is trivial,
 as a primary work for the general cases with non-trivial holonomy.
The role of each parameter will be clarified in the following sections.

%%%%%%%%%%%%%%%%%%%%%%%%%%%%%%%%%%%%%%%%%%%%%%%%%%%%%%%%%%%%%%%

We should mention the relationship between the earlier works on the Nahm transform of calorons and the present work.
In \cite{Bruckmann:2004nu}, the authors considered the Nahm transform by utilizing the Green function method.
In contrast, we concentrate here on the Nahm transform by solving numerically the Weyl equations with impurities.
These two ways lead, of course, equivalent results for various physical quantities such as action density of field configuration.
However, we expect that the solutions to the Weyl equations in the presence of impurities and their numerical code considered here
 make a crucial contribution to the progress of Nahm transform for diverse objects \cite{Jardim},
 and also D-brane theories, \textit{e.g.}, \cite{Kapustin-Sethi}. 

This Letter is organized as follows.
In section 2, we give a brief review on the Nahm transform of calorons.
In section 3, we make a formulation for numerical analysis to the Weyl equations with impurities.
In section 4, we consider the action density of 2-caloron and give the interpretation of moduli parameters.
Section 5 is devoted for concluding remarks.

%%%%%%%%%%%%%%%%%%%%%%%%%%%%%%%%%%%%%%%%%%%%%%%%%%%%%%%%%%%%%%%%%%
%\setcounter{section}{1}

\section{Nahm construction for {\it k}-calorons}
\subsection{Basic formalism}
%%%%%%%%%%%%%%%%%%%%%%%%%%%%%%%%%%%%%%%%%%%%%%%%%%%%%%%%%%%%%%%%%%%%%%

In this subsection, we give a brief review of the Nahm transform for SU(2) $k$-caloron with trivial holonomy.
As mentioned in Introduction, the Nahm data of calorons consists of the bulk Nahm data and the boundary Nahm data.
The bulk one is $k\times k$ Hermite matrices  $T_\mu(s)$ periodic in $s$ with period $\mu_0$, where $\mu=0,1,2,3$ and $s\in I$.
They are smooth functions except for at the boundaries, with fundamental interval, say, $I = (- \mu_0 / 2, \mu_0 / 2)$.
For the gauge field to be ASD, they are subject to the Nahm equations,
\begin{align} \label{eqn:the bulk Nahm equation} %\ref{eqn:the bulk Nahm equation}
    & \frac{\mathrm d}{\mathrm ds}T_i(s) - \mathrm i\big[T_0(s), T_i(s)\big] - \frac{\mathrm i}{2}\varepsilon_{ijk}\big[T_j(s), T_k(s)\big]
    = 0,
\end{align}
where the roman subscripts are $1,2$ or $3$,
together with the reality conditions $T_\mu(- s) = \!\,^\mathrm tT_\mu (s)$.
On the other hand, the boundary Nahm data is given by a $k$-row vector $W$ of quaternion entries, enjoying 
the matching conditions
\begin{align} \label{eqn:the matching condition} %\ref{eqn:the matching condition}
    & T_j(- \mu_0 / 2) - T_j(\mu_0 / 2)
    = \frac{1}{2}\mathrm{Tr}_2( W^\dagger W\sigma_j ),
\end{align}
where $\sigma_j$'s are the Pauli matrices and  the trace is taken over quaternions.

The link between the Nahm construction of calorons and that of monopoles is as follows.
In the construction of monopoles, the Nahm data should also enjoy (\ref{eqn:the bulk Nahm equation})
and have simple poles at the boundary of the interval
\cite{Hitchin, Houghton:1995bs} to give correct asymptotic forms of the Higgs fields.
We can make use of their ``non-singular parts" as the bulk Nahm data of calorons \cite{Ward:2003sx}, 
which are piece of the monopole data attached with the shorter segment of the interval.
If we take the monopole limit of the calorons, the interval $I$ eventually fills the characteristic period of the bulk data so that
the simple poles appear at the boundaries, accordingly the matching conditions erupt in certain cases.

%%%%%%%%%%%%%%%%%%%%%%%%%%%%%%%%%%%%%%%%%%%%%%%%%%%%%%%%%%%%%%%

The caloron gauge fields in the real, or configuration space are obtained from the Nahm data through the zero modes of 
the Weyl equations with impurities at the boundaries. 
Denoting a ``spinor" $U(s;x^\alpha)$ defined on $I$ with period $\mu_0$ 
and a single quaternion $V(x^\alpha)$, the Weyl equations are
\begin{align} \label{eqn:the Weyl equation} %\ref{eqn:the Weyl equation}
    & \big\{\bm1_{2k}\frac{\mathrm d}{\mathrm ds} - \mathrm i\big(T_\mu(s) + x_\mu\bm1_k\big) \otimes e_\mu\big\}U(s;x^\alpha)
      \nonumber \\ & \qquad \qquad \qquad \qquad \qquad
    = \mathrm iW^\dagger V(x^\alpha)\delta(s - \mu_0 / 2),
\end{align}
 where $\bm1_k$ is the $k \times k$ unit matrix
 and $e_\mu = (\bm1_2, - \mathrm i\sigma_j)$ are the basis of the quaternion.
We write these zero modes into components as
\begin{align}
    & U(s;x^\alpha)
    = (\bm u_1, \bm u_2), \quad
      V(x^\alpha)
    = (\bm v_1, \bm v_2),
\end{align}
where $\bm u_1$, $\bm u_2$ and $\bm v_1$, $\bm v_2$ 
are $2k$-column vectors and $2$-column vectors, respectively.
By using them, we can separate (\ref{eqn:the Weyl equation}) into two equations, the bulk Weyl equations on $I$
\begin{align} \label{eqn:bulk Weyl equation} %\ref{eqn:bulk Weyl equation}
    & \big\{\bm1_{2k}\frac{\mathrm d}{\mathrm ds}
 - \mathrm i\big(T_\mu(s) + x_\mu\bm1_k\big) \otimes e_\mu\big\}\bm u_\ell
    = 0,
\end{align}
and the matching conditions at the boundary
\begin{align} \label{eqn:boundary Weyl equation} %\ref{eqn:boundary Weyl equation}
    & \Delta\bm u_\ell
    \equiv \bm u_\ell(- \mu_0 / 2) - \bm u_\ell(\mu_0 / 2)
    = \mathrm iW^\dagger \bm v_\ell,
\end{align}
where $\ell = 1, 2$.
The next step is to find two independent pair of the zero modes $(\bm u_1, \bm v_1)$, $(\bm u_2, \bm v_2)$,
orthonormalized  as
\begin{align} \label{eqn:orthonormalization} %\ref{eqn:orthonormalization}
    & \int_I\bm u_a^\dagger\bm u_b\mathrm ds
                    + \bm v_a^\dagger\bm v_b
    = \delta_{ab},
\end{align}
where $a, b = 1, 2$.
Putting together these zero modes, we obtain the ASD gauge connection of calorons as
\begin{align} \label{eqn:the Nahm construction} %\ref{eqn:the Nahm construction}
    & \{A_\alpha(x)\}_{ab}
    = \int_I\bm u_a^\dagger\partial_\alpha\bm u_b\mathrm ds
                    + \bm v_a^\dagger\partial_\alpha\bm v_b.
\end{align}
As mentioned in Introduction, we have to perform the Nahm transform by numerical analysis, 
which is the main aim of this Letter.
Note that the numerical Nahm transform for monopoles was vigorously studied by Houghton and Sutcliffe
 \cite{Houghton:1995bs}.

\bigskip
%%%%%%%%%%%%%%%%%%%%%%%%%%%%%%%%%%%%%%%%%%%%%%%%%%%%%%%%%%%%%%%

\subsection{Exact Nahm data of the 2-caloron}

In this subsection, we introduce an exact Nahm data of the 2-caloron, for which we perform the Nahm transform in the following
sections.
In \cite{Nakamula:2009sq}, the bulk Nahm data of the 2-caloron on an interval is given in the following form
\begin{align} \label{eqn:2-caloron bulk data} %\ref{eqn:2-caloron bulk data}
      T_1(s) &
    = f_1(s)\sigma_1~~~~~~~~~~~+ g_1(s)\sigma_3 + d_1\bm1_2, \nonumber \\
      T_2(s) &
    =               ~~~~~~~~~~\,
      f_2(s)\sigma_2~~~~~~~~~~~~~\,+ d_2\bm1_2, \nonumber \\
      T_3(s) &
    = g_3(s)\sigma_1~~~~~~~~~~~+ f_3(s)\sigma_3 + d_3\bm1_2, \nonumber \\
      T_0~~~~&
    = ~~~~~~~~~~~~~~~~~~~~~~~~~~~~~~~~~~~~~\,d_0\bm1_2.
\end{align}
The solutions to the Nahm equations (\ref{eqn:the bulk Nahm equation}) are
\begin{align}\label{eqn:bulk solutions}
    &   f_1(s)
    =   ~~a(s)\cos\phi, \qquad \quad
        g_1(s)
    =   - b(s)\sin\phi,\nonumber  \\
    & ~~~~~~~~~~~~~~~~~f_2(s)
    = \mp Dk'\frac{\mathrm{sn}2Ds}{\mathrm{cn}2Ds},\nonumber  \\
    &   g_3(s)
    = \pm a(s)\sin\phi, \qquad \quad
        f_3(s)
    = \pm b(s)\cos\phi,
\end{align}
 with
\begin{align} \label{eqn:coefficients of the bulk data}
    & \big(a(s),b(s)\big)
    = \left(D\frac{            k'}{\mathrm{cn}2Ds}
           ,D\frac{\mathrm{dn}2Ds}{\mathrm{cn}2Ds}\right)
      \nonumber \\ &
      \qquad \qquad \quad \mathrm{or}\,
      \left(D\frac{\mathrm{dn}2Ds}{\mathrm{cn}2Ds}
           ,D\frac{            k'}{\mathrm{cn}2Ds}\right),
\end{align}
where $\mathrm{sn}$, $\mathrm{cn}$, $\mathrm{dn}$ are  Jacobi elliptic functions 
of modulus $k$, and $k' = \sqrt{1 - k^2}$.
The monopole limit of this bulk data is obtained by putting $\mu_0/2\to1$ and $D\to K(k)/2$.
Note that  $\phi$ is not a physical parameter for the trivial holonomy calorons and also monopoles,
because it can be removed by a spatial rotation (see section IV).
However, we keep it for the purpose of generalization to the non-trivial holonomy cases, 
and also the reliability check of the numerical code.

%%%%%%%%%%%%%%%%%%%%%%%%%%%%%%%%%%%%%%%%%%%%%%%%%%%%%%%%%%%%

%\subsection{The matching condition}
%%%%%%%%%%%%%%%%%%%%%%%%%%%%%%%%%%%%%%%%%%%%%%%%%%%%%%%%%%%%%%%%%

Next, we consider the matching condition (\ref{eqn:the matching condition}) for the 2-caloron.
In accordance with  \cite{Nakamula:2009sq}, we employ the following parameterizations for the boundary
Nahm data
\begin{align}\label{eqn:matching condition}
    & W
    = (\lambda\bm1_2,\rho\hat q), \quad
      \hat q
    = \hat q_\mu e_\mu, \quad
      \hat q_0
    = \cos\psi,
      \nonumber \\ &
      \hat{\bm q}
    = (\sin\psi\sin\theta\sin\varphi, \sin\psi\sin\theta\cos\varphi, \sin\psi\cos\theta).
\end{align}
Then (\ref{eqn:the matching condition})  reads
\begin{align}\label{eqn:matching condition2}
    & T_j(- \mu_0 / 2) - T_j(\mu_0 / 2)
    = \lambda\rho\hat q_j\sigma_2.
\end{align}
From the boundary values of the bulk Nahm data given above,
we find  $\theta = \pi / 2$, $\varphi = 0~\mathrm{or}~\pi$, and 
\begin{align} \label{eqn:remaining matching condition} %\ref{eqn:remaining matching condition}
    & \pm 2Dk'\frac{\mathrm{sn}2D(\mu_0 / 2)}{\mathrm{cn}2D(\mu_0 / 2)}
    = \lambda\rho\sin\psi.
\end{align}
Note that all the parameters are assumed to be real number.
Consequently, the moduli parameters of the 2-caloron of trivial holonomy are the following 10,
\begin{align} \label{eqn:moduli parameters} %\ref{eqn:moduli parameters}
    & d_1, d_2, d_3, d_0, \phi, k, D, \lambda, \rho, \psi,
\end{align}
which are subject to the constraint (\ref{eqn:remaining matching condition}).

\bigskip
%%%%%%%%%%%%%%%%%%%%%%%%%%%%%%%%%%%%%%%%%%%%%%%%%%%%%%%%%%%

\begin{figure*}[tbp]
\centering
\includegraphics[clip,width=16cm,height=16cm,keepaspectratio]{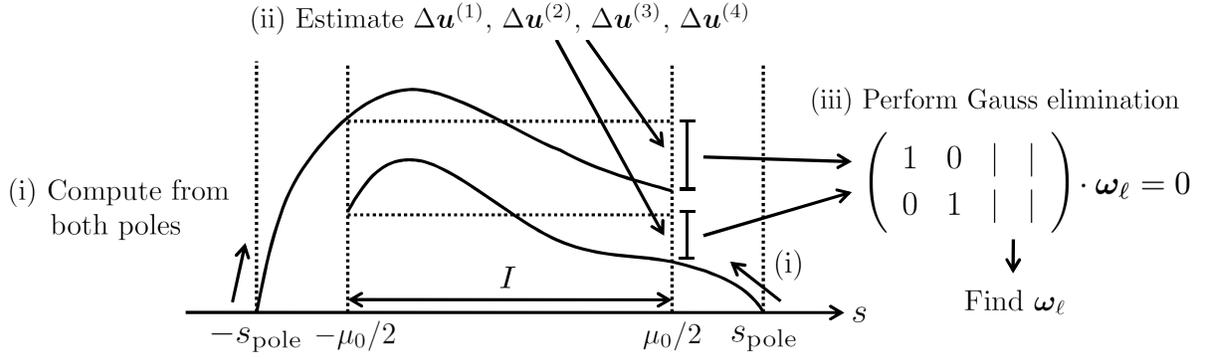}
\caption{
The method of solving the boundary Weyl equation removing the over-determination.
We compute the bulk Weyl equation from both poles independently.
Then four of them are used to determine linear combinations which can be performed with the Gauss elimination for the coefficient 
matrix of (\ref{eqn:conditions to remove overdetermined 2}).
} \label{fig:solve_Weyl_eqn} %\ref{fig:solve_Weyl_eqn}
\end{figure*}

%%%%%%%%%%%%%%%%%%%%%%%%%%%%%%%%%%%%%%%%%%%%%%%%%%%%%%%%%%%%

\section{Numerical  Nahm transform: Formulation}

\subsection{Solving the bulk Weyl equation}
%%%%%%%%%%%%%%%%%%%%%%%%%%%%%%%%%%%%%%%%%%%%%%%%%%%%%%%%%%%

In this section, we give the strategy on finding out the numerical solution to the Weyl equations with boundary impurities,
which is based on the construction of monopoles \textit{a la} Houghton and Sutcliffe \cite{Houghton:1995bs}.
From the Nahm data (\ref{eqn:2-caloron bulk data}),
the bulk Weyl equations (\ref{eqn:the Weyl equation}) can be written 
for the components of $\bm u_\ell \equiv \!\,^\mathrm t(u_{1\ell}, u_{2\ell}, u_{3\ell}, u_{4\ell})$ as

\begin{widetext}
\begin{align} \label{eqn:decomposed bulk Weyl equation} %\ref{eqn:decomposed bulk Weyl equation}
    & \left\{
          \left(\begin{array}{cccc}
              1 &   &   & O \\
                & 1 &   &   \\
                &   & 1 &   \\
              O &   &   & 1
          \end{array}\right)
          \frac{\mathrm d}{\mathrm ds}
        - \left(\begin{array}{cccc}
                    f_3 &       g_1 &       g_3 & f_1 - f_2 \\
                    g_1 &     - f_3 & f_1 + f_2 &     - g_3 \\
                    g_3 & f_1 + f_2 &     - f_3 &     - g_1 \\
              f_1 - f_2 &     - g_3 &     - g_1 &       f_3
          \end{array}\right)
      \right.
    - \left(\begin{array}{cccc}
          d_3 + \mathrm id_0 &   d_1 - \mathrm id_2 &                  0 &                    0 \\
          d_1 + \mathrm id_2 & - d_3 + \mathrm id_0 &                  0 &                    0 \\
                           0 &                    0 & d_3 + \mathrm id_0 &   d_1 - \mathrm id_2 \\
                           0 &                    0 & d_1 + \mathrm id_2 & - d_3 + \mathrm id_0
      \end{array}\right) \nonumber \\ &
      \left.
        - \left(\begin{array}{cccc}
              x_3 + \mathrm ix_0 &   x_1 - \mathrm ix_2 &                  0 &                    0 \\
              x_1 + \mathrm ix_2 & - x_3 + \mathrm ix_0 &                  0 &                    0 \\
                               0 &                    0 & x_3 + \mathrm ix_0 &   x_1 - \mathrm ix_2 \\
                               0 &                    0 & x_1 + \mathrm ix_2 & - x_3 + \mathrm ix_0 \end{array}\right)
      \right\}
      \left(\begin{array}{c} u_{1\ell} \\ u_{2\ell} \\ u_{3\ell} \\ u_{4\ell} \end{array}\right)
    = 0.
\end{align}
\end{widetext}
This system of ordinary differential equations can be solved by the Runge-Kutta method with appropriate initial conditions.
For the Nahm transform of calorons, as well as monopoles, 
 what we need is the basis of 2-dimensional vector space spanned by the solutions to (\ref{eqn:decomposed bulk Weyl equation}) 
 normalizable on $I$.
To employ the monopole construction procedure, we have to integrate (\ref{eqn:decomposed bulk Weyl equation})
starting from the simple poles of the Nahm data.
The positions of poles are $s_{\mathrm{pole}}=\pm K(k)/2D$, obviously read from (\ref{eqn:bulk solutions}) and
(\ref{eqn:coefficients of the bulk data}).
From the Nahm construction of calorons,
  $s_{\mathrm{pole}}$'s are assumed to be located on the outside of $I$,  \textit{i.e.}, $|s_{\mathrm{pole}}|>\mu_0/2$.
If we represent (\ref{eqn:decomposed bulk Weyl equation}) at each pole in the following concise form  
\begin{align}
(s-s_\mathrm{pole})\frac{\mathrm d\bm u_\ell}{\mathrm ds} = B_s\bm u_\ell,
\end{align}
where $B_s$ is a regular matrix, 
then the vector space dimensions of independent solutions which are normalizable 
are given by the number of positive eigenvalues of the matrix $B_{s_\mathrm{pole}}$ \cite{Houghton:1995bs}.
In the present case, we evaluate that the eigenvalues at each pole are, respectively,
\begin{eqnarray}
\frac{1}{2},~~\frac{1}{2},~~\frac{-1\pm 2\sqrt{k + 1}}{2}  
\end{eqnarray}
for an arbitrary $\phi$.
Hence, we find the vector space dimensions are three for each pole.
In terms of Frobenius expansion at the poles, we can find a couple of free parameters.
Appropriately tuning these parameters, we obtain the three independent initial conditions at each pole.  
In this way, we can compute the array of  independent solutions $\bm{u}^{(1)},\bm{u}^{(2)},\cdots,\bm{u}^{(n)}$, 
where $n$ is at most 6.

The 2-dimensional basis $\bm{u}_\ell$, $(\ell=1,2)$, can be obtained by taking linear combinations of these solutions,
\begin{eqnarray}\label{eqn:linearcombination}
\bm{u}_\ell=(\bm{u}^{(1)},\bm{u}^{(2)},\cdots,\bm{u}^{(n)})\cdot \bm{\omega}_\ell\,,
\end{eqnarray}
where $\bm{\omega}_\ell$ are  $n$-column vectors,
\textit{i.e.}, $\bm{\omega}_\ell:=^{\rm t}(\omega_{1\ell},\omega_{2\ell},\cdots,\omega_{n\ell})$,
to be fixed below.
The evaluation of $\bm{\omega}_\ell$ is not so straightforward for calorons, as well as monopoles. 
For the monopole construction \cite{Houghton:1995bs}, $\bm{\omega}_\ell$ is determined by 
the condition that a linear combination of the solutions, integrated from one pole, matches 
a linear combination of the solutions from the other pole, at the center of the interval.
In this way, we find the 2-dimensional basis of the vector space.
For the caloron construction, the situation is slightly different.
We are able to compute the solutions in the whole interval at once without taking linear combination.  
Instead, we have to find out the solutions to
 the boundary Weyl equations (\ref{eqn:boundary Weyl equation}) simultaneously, 
which should be consistent with the boundary Nahm data.

\subsection{Solving the boundary Weyl equation}

For an appropriately defined $\bm{u}_\ell$, the boundary Weyl equations (\ref{eqn:boundary Weyl equation}) 
with the boundary Nahm data (\ref{eqn:matching condition}) are 
\begin{align} \label{eqn:boundary Weyl equation2} %\ref{eqn:boundary Weyl equation2}
    & \left(\begin{array}{c} \Delta u_{1\ell} \\ \Delta u_{2\ell} \\ \Delta u_{3\ell} \\ \Delta u_{4\ell} \end{array}\right)
    = \mathrm i
      \left(\begin{array}{cc}
          ~~     \lambda &            0 \\
          ~~           0 &      \lambda \\
          ~~\rho\cos\psi & \rho\sin\psi \\
          - \rho\sin\psi & \rho\cos\psi
      \end{array}\right)
      \left(\begin{array}{c} v_{1\ell} \\ v_{2\ell} \end{array}\right),
\end{align}
 where
 $\Delta\bm u_\ell = \!\,^\mathrm t(\Delta u_{1\ell}, \Delta u_{2\ell}, \Delta u_{3\ell}, \Delta u_{4\ell})$
 and
 $\bm v_\ell = \!\,^\mathrm t(v_{1\ell}, v_{2\ell})$.
The equations seem to be an {\it over-determined system} for $v_{1\ell},v_{2\ell}$, 
but that is not true because the left-hand side is not determined at this stage.
Our goal is to determine the coefficients of the linear combination $\bm{\omega}_\ell$ in (\ref{eqn:linearcombination}) and  $\bm v_\ell$, simultaneously, 
for the given bulk Nahm data $\bm{u}^{(1)},\bm{u}^{(2)},\cdots,\bm{u}^{(n)}$.
This procedure can be reduced to the problem of linear algebra as follows.

First, we solve the upper two rows of (\ref{eqn:boundary Weyl equation2}) as
\begin{align} \label{eqn:the solutions of the boundary Weyl equation} %\ref{eqn:the solutions of the boundary Weyl equation}
    & v_{\ell1}
    = - \mathrm i\Delta u_{\ell1} / \lambda, \quad
      v_{\ell2}
    = - \mathrm i\Delta u_{\ell2} / \lambda\,.
\end{align}
By substituting (\ref{eqn:the solutions of the boundary Weyl equation}) into the lower two rows, we find
 (\ref{eqn:boundary Weyl equation2}) becomes the constraints on   $\Delta\bm u_\ell$ as
\begin{align} \label{eqn:conditions to remove overdetermined} %\ref{eqn:conditions to remove overdetermined}
      \Delta u_{\ell3} &
    = \rho\big(\Delta u_{\ell1}\cos\psi + \Delta u_{\ell2}\sin\psi\big) / \lambda,
      \nonumber \\
      \Delta u_{\ell4} &
    = \rho\big(\Delta u_{\ell2}\cos\psi - \Delta u_{\ell1}\sin\psi\big) / \lambda.
\end{align}
The next step is to fix $\bm{\omega}_\ell$ which is consistent with (\ref{eqn:conditions to remove overdetermined}).
From the definition (\ref{eqn:linearcombination}),  $\Delta \bm{u}_\ell$ is expanded as
\begin{align} \label{eqn:linear combination of L.H.S.} %\ref{eqn:linear combination of L.H.S.}
    & \Delta\bm u_\ell
    = (\Delta\bm u^{(1)}, \Delta\bm u^{(2)}, \cdots, \Delta\bm u^{(n)}) \cdot \bm\omega_\ell,
\end{align}
 where $\Delta\bm u^{(i)} \equiv \bm u^{(i)}(- \mu_0 / 2) - \bm u^{(i)}(\mu_0 / 2)$.
Note that $\Delta\bm u_\ell$ and $\Delta\bm u^{(i)}_\ell$'s are 4-column vectors.
Now, we can rearrange   (\ref{eqn:conditions to remove overdetermined}), by using (\ref{eqn:linear combination of L.H.S.}),
into a linear equation for $\Delta\bm u_\ell$ 
\begin{align} \label{eqn:conditions to remove overdetermined 2} %\ref{eqn:conditions to remove overdetermined 2}
    & \left(\begin{array}{cccc}
          ~~\rho\cos\psi & \rho\sin\psi &     - \lambda & \quad       0 \\
          - \rho\sin\psi & \rho\cos\psi & \quad       0 &     - \lambda
      \end{array}\right)
      \nonumber \\ & \qquad \qquad
      \cdot (\Delta\bm u^{(1)}, \Delta\bm u^{(2)}, \cdots, \Delta\bm u^{(n)}) \cdot \bm\omega_\ell
    = 0.
\end{align}
This can also be regarded as the linear defining equation for $\bm\omega_\ell$.
Since the number  of constraints on  $\bm\omega_\ell$ in  (\ref{eqn:conditions to remove overdetermined 2}) is two,
 we have to take at least four solutions of the bulk Weyl equation, to obtain two independent solutions  $\bm\omega_\ell$,
\textit{i.e.}, the 2-dimensional basis  $\bm{u}_\ell$.
To determine the independent solutions $\bm{\omega}_{\ell}$ to (\ref{eqn:conditions to remove overdetermined 2}),
we simply perform the Gauss elimination (Fig.\ref{fig:solve_Weyl_eqn}).
Having obtained the independent  $\bm{\omega}_{\ell}$, we  take the linear combination (\ref{eqn:linear combination of L.H.S.}),
and find the components of $\Delta\bm u_\ell$ automatically satisfy (\ref{eqn:conditions to remove overdetermined}).
The solutions of the boundary Weyl equation have already been given by (\ref{eqn:the solutions of the boundary Weyl equation}).
Our method to solve the Weyl equations is schematically illustrated in Fig.\ref{fig:solve_Weyl_eqn}.

%%%%%%%%%%%%%%%%%%%%%%%%%%%%%%%%%%%%%%%%%%%%%%%%%%%%%%%%%%%%

\begin{figure*}
\centering
\includegraphics[clip,width=5cm,height=5cm,keepaspectratio]{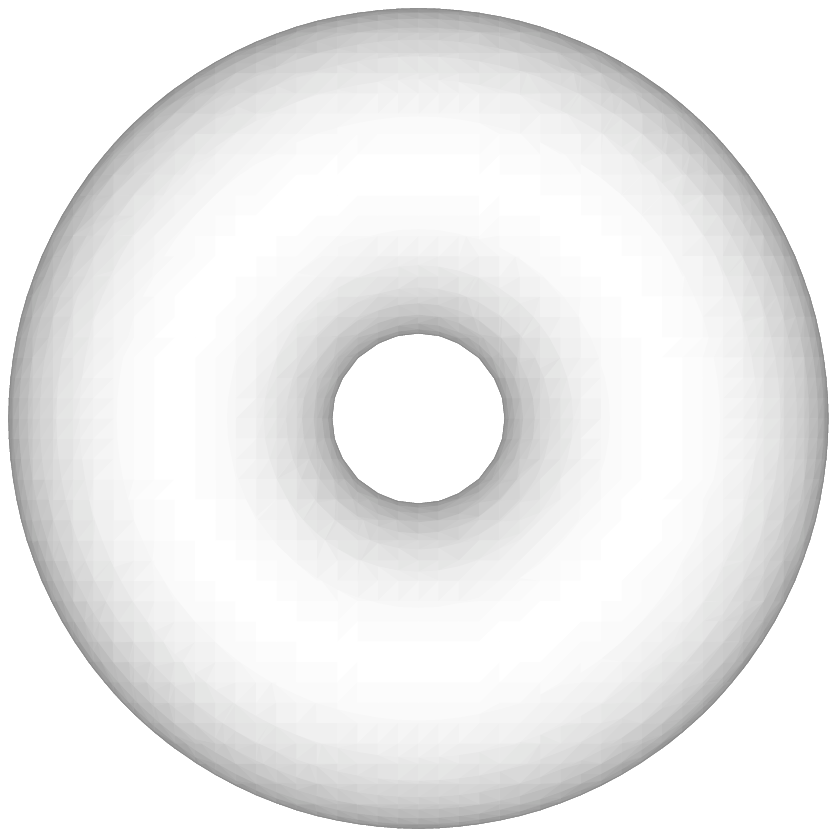}~~~~
\includegraphics[clip,width=5cm,height=5cm,keepaspectratio]{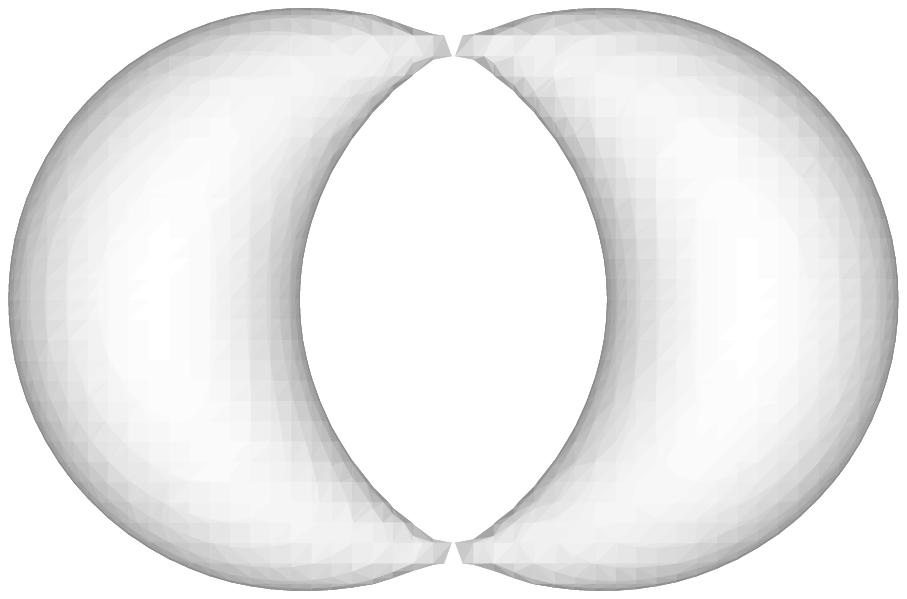}~~~~
\includegraphics[clip,width=5cm,height=5cm,keepaspectratio]{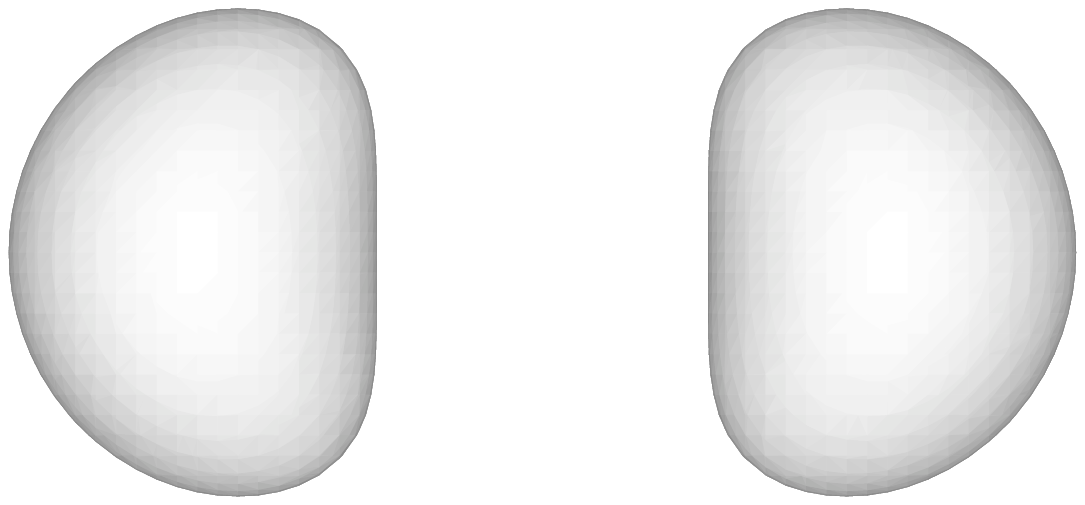}
\caption{
Action isosurface plot of the 2-caloron with $\mu_0 / 2 = 0.5$, $t = 0$, $S(\bm x, t) = 2.94$, $k = 0.0, 0.7, 0.9$ respectively.
} \label{fig:isoaction_plot} %\ref{fig:isoaction_plot}
\end{figure*}

\subsection{Realization of the action density}

To investigate the configuration of calorons in the configuration space,
 it is useful to visualize the action density, which is gauge invariant and positive real definite.
With the fact that the field strength of calorons
 $F_{\mu\nu} = \partial_\mu A_\nu - \partial_\nu A_\mu + [A_\mu, A_\nu]$
 is an anti-symmetric tensor and the (A)SD conditions
 $F_{01} = \pm F_{32}$, $F_{02} = \pm F_{31}$, $F_{03} = \pm F_{21}$,
 the action density of calorons can be written as
\begin{align} \label{eqn:the action density} %\ref{eqn:the action density}
    & S(\bm x, t)
    = - \textstyle\frac{1}{2}\mathrm{Tr}_2F_{\mu\nu}^2
    = - 2\mathrm{Tr}_2(F_{12}^2 + F_{23}^2 + F_{31}^2).
\end{align}
Thanks  to the (A)SD conditions, 
 we do not have to calculate the ``time" $t = x_0$ derivative in the field strength
so that we can regard $t$ as a parameter in the calculation.
We perform the visualization of the action density of the caloron, mainly 
by Mathematica~\cite{wolfram}.

\section{Numerical Nahm transform: Analysis} \label{Results of the comptation} %\ref{Results of the comptation}

\subsection{The patchwork}

Combining all these procedures, we now compute the action density of the calorons, a typical gauge invariant quantity.
However, when we carry out the program, we always observe unexpected singularities of line form.  
In the calculation of the zero modes of the Weyl equation,
 we always have a trivial phase factor, which depends on the configuration space coordinates. 
These factors inevitably cause effect to the gauge field through a finite difference method of (\ref{eqn:the Nahm construction}). 
For the analysis of the monopoles \cite{Houghton:1995bs}, on the contrary, no such singularities occur. 
The finite difference of the zero modes is taken after the trace for quaternion, 
which successfully cancels jump or twist of the phase of the zero modes.

These singularities, however, have no essential ingredients so we employ the following simple procedure to remove them.
We can identify numerically the location of the singular lines, which depends on the initial conditions of the Runge-Kutta method. 
For example, a solution has a singular line located on the upper hemisphere of 
the configuration space while another has it on the lower hemisphere.
Thus, we obtain the regular action density by a patchwork with two or three parts of the solutions
with different parameter sets, by choosing their intersection is regular.
We employ 50 grid points for the dual space, and $61 \times 61 \times 61$ lattice points for the configuration space, 
which are sufficient for the numerical convergence. 

\begin{figure*}
\begin{minipage}{7cm}
\centering
\includegraphics[clip,width=7cm,height=7cm,keepaspectratio]{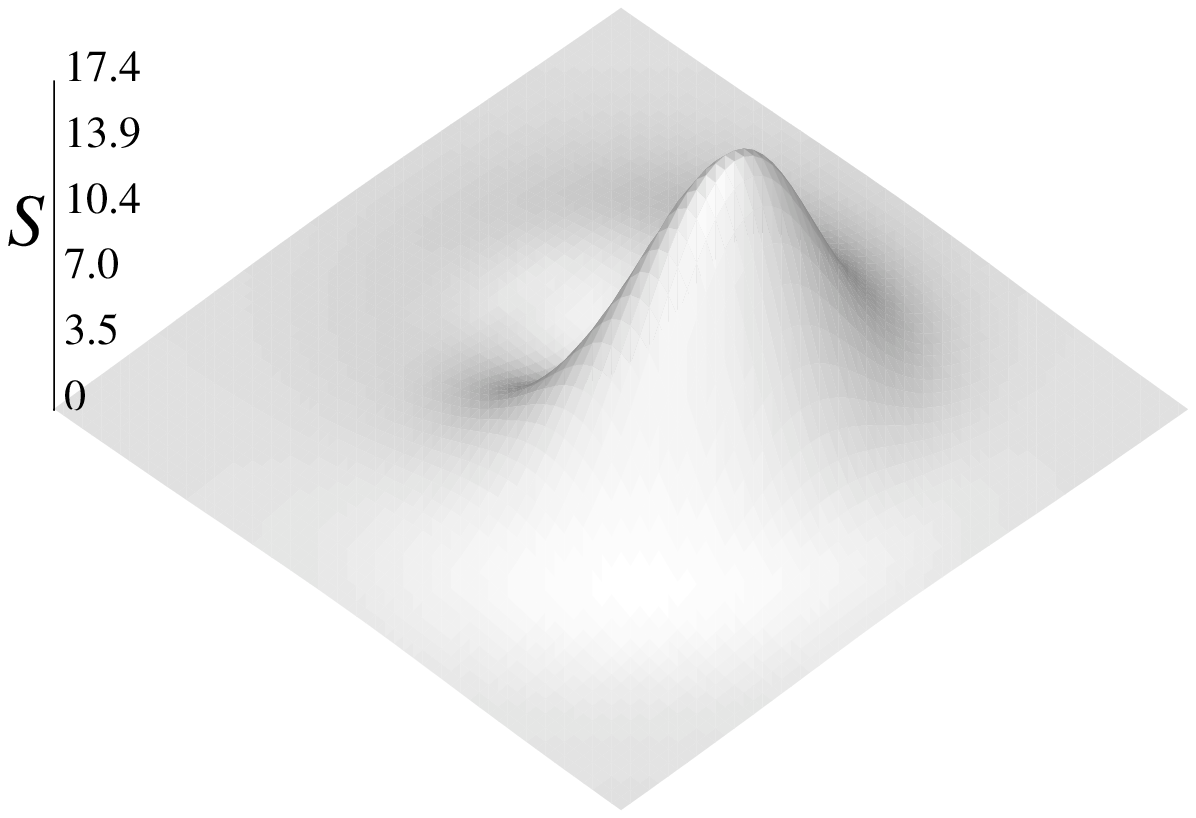} \\
\vspace{5mm}

\includegraphics[clip,width=7cm,height=7cm,keepaspectratio]{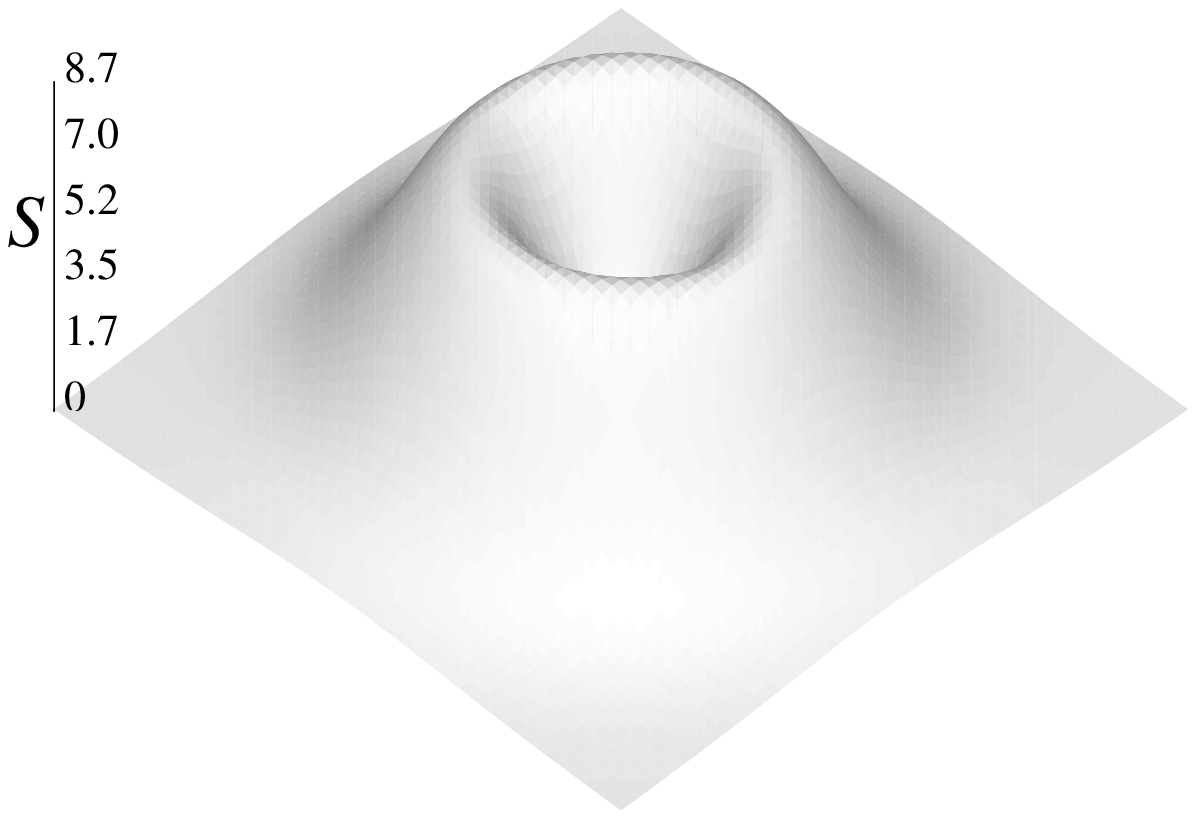} \\
\vspace{5mm}

\includegraphics[clip,width=7cm,height=7cm,keepaspectratio]{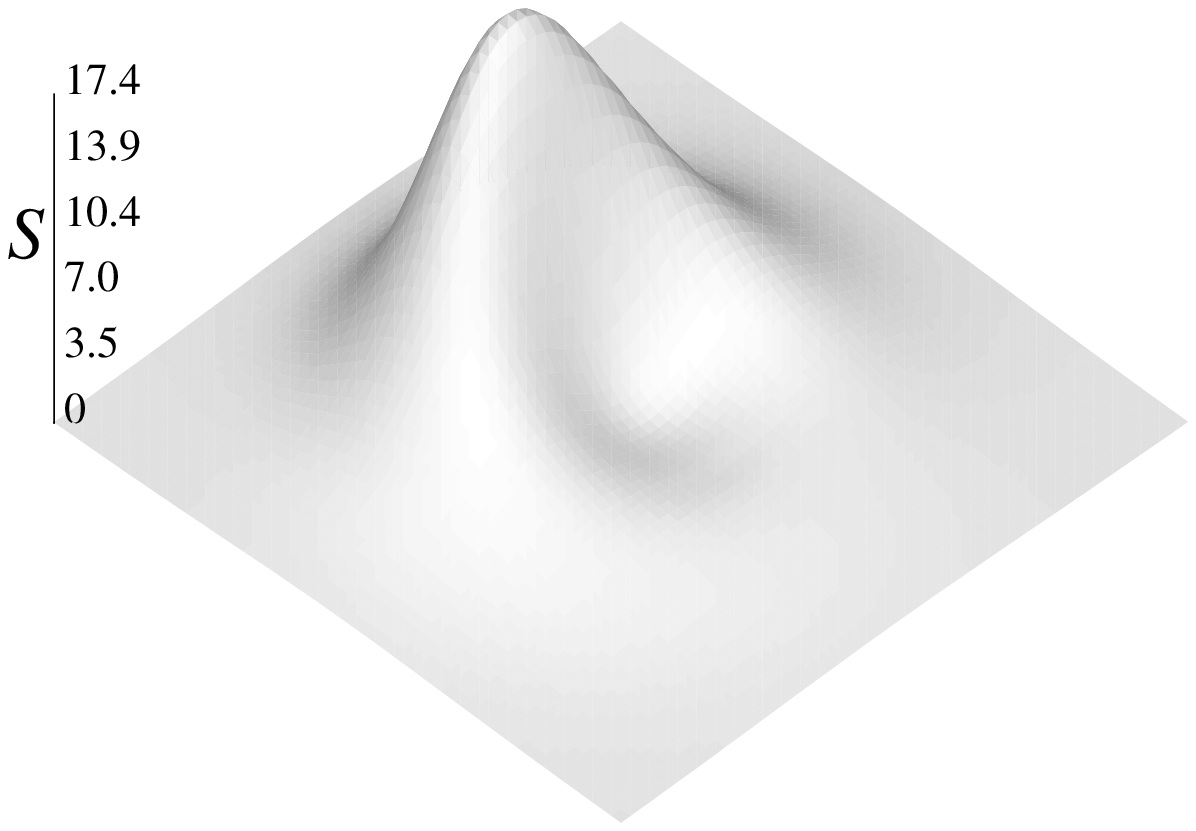}
\caption{
The action densities on $x_1x_3$-plane with $\mu_0 / 2 = 0.5$,
 $t = 0$, $\psi = \pi / 2$, for $\gamma :=\lambda/\rho= 3$~(top), $\gamma=1$~(middle), $\gamma=1/3$~(bottom), respectively.
} \label{fig:action_density_psi_rho1} %\ref{fig:action_density_psi_rho1}
\end{minipage}~~~~~~
\begin{minipage}{7cm}
\centering
\includegraphics[clip,width=7cm,height=7cm,keepaspectratio]{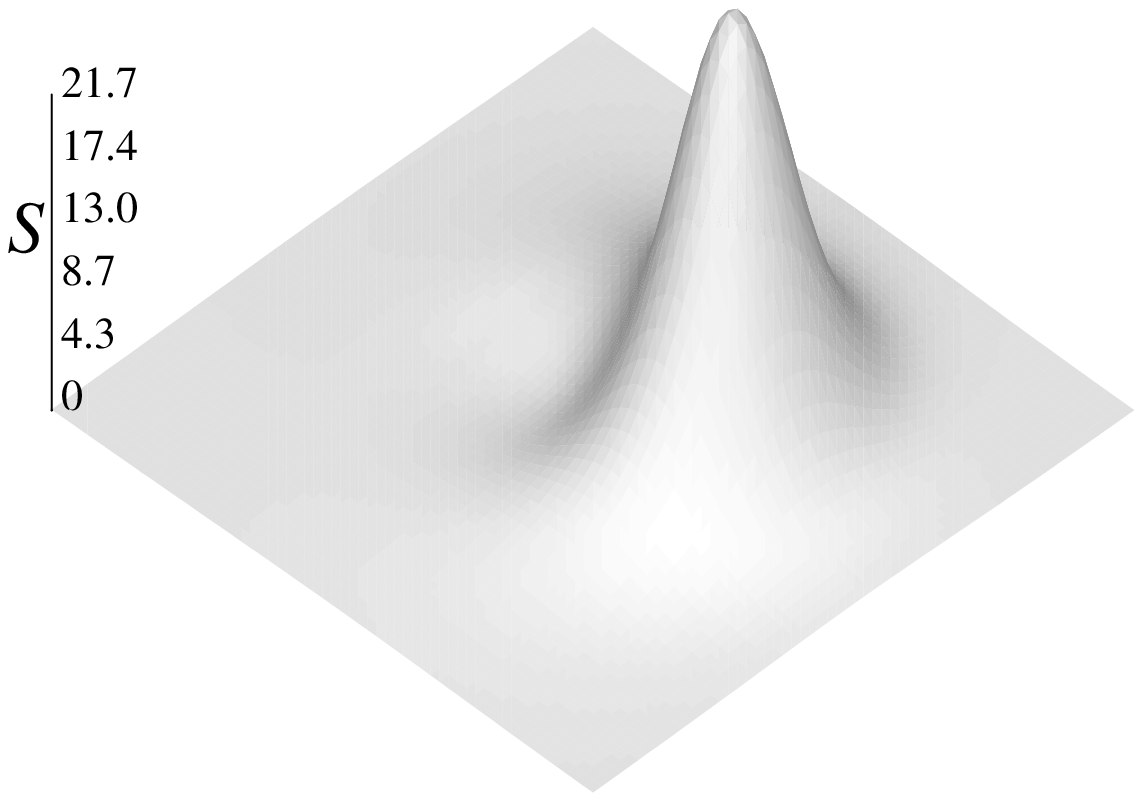} \\
\vspace{5mm}

\includegraphics[clip,width=7cm,height=7cm,keepaspectratio]{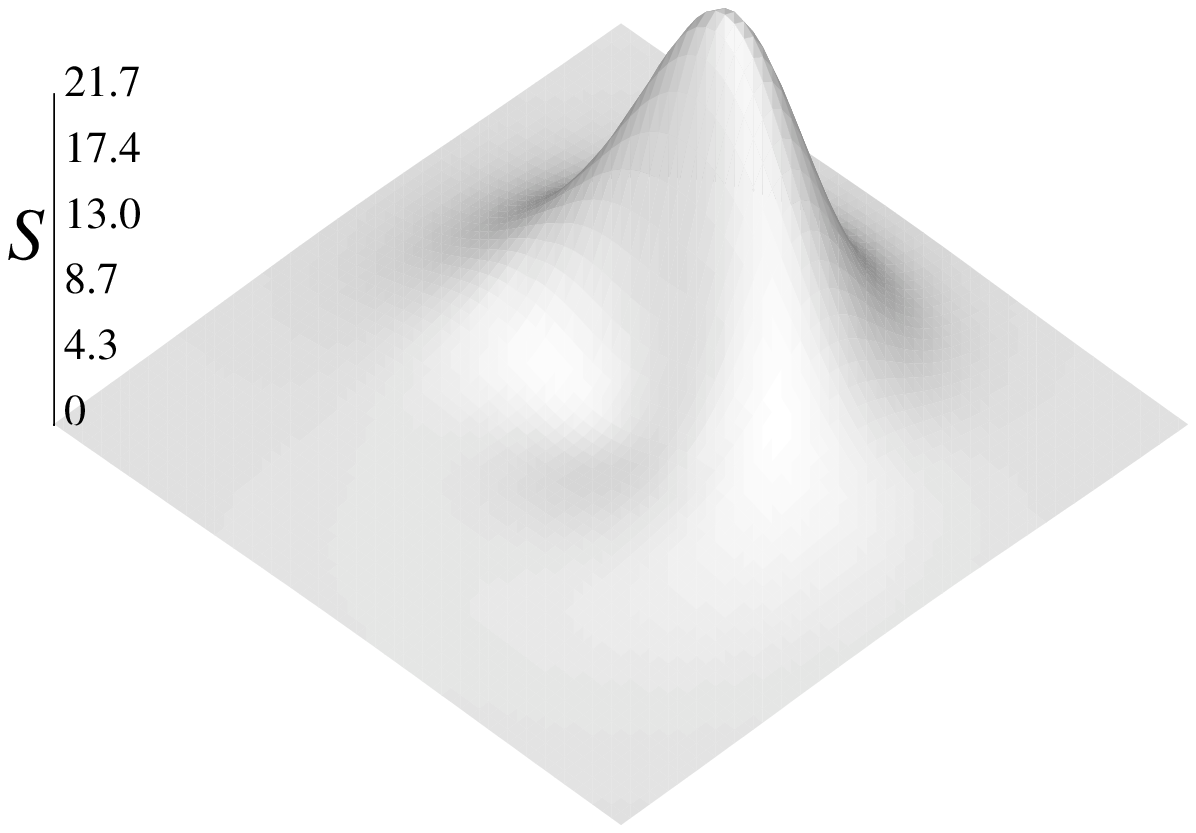} \\
\vspace{5mm}

\includegraphics[clip,width=7cm,height=7cm,keepaspectratio]{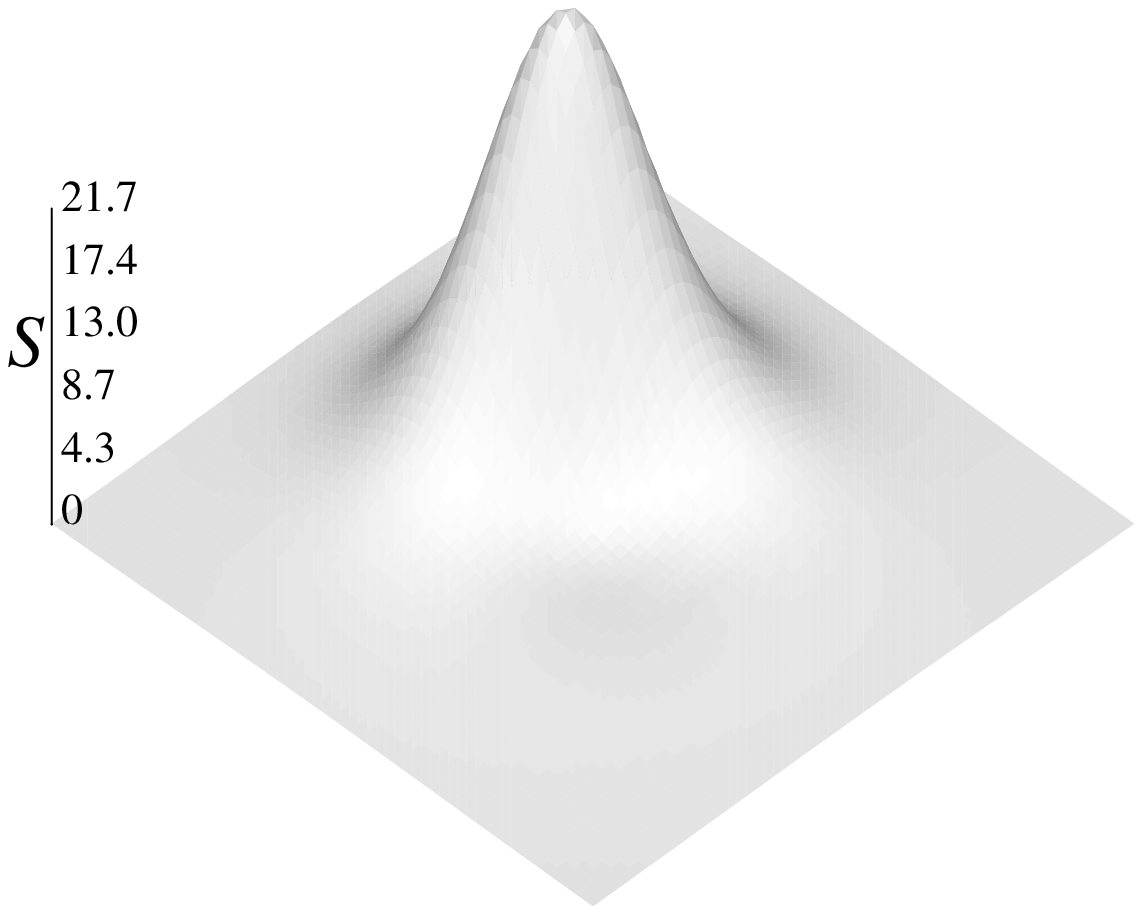}
\caption{
Similar with the Fig.\ref{fig:action_density_psi_rho1} but with the case of 
$\psi=3\pi/20$.
} 
\label{fig:action_density_psi_rho2} %\ref{fig:action_density_psi_rho2}
\end{minipage}
\end{figure*}

\subsection{The moduli parameters}

Having established the formulation for calculating the action density,
our main concern is to clarify the role of the moduli parameters of the 2-calorons, 
which are listed in (\ref{eqn:moduli parameters}).

First, we consider a very simple case
\begin{align} \label{eqn:some replacements} %\ref{eqn:some replacements}
    & d_1
    = d_2
    = d_3
    = d_0
    = 0, \quad
      k
    = 0,
      \nonumber \\ &
      D
    = K(k) / 2
    = \pi / 4, \quad
      \rho
    = \lambda, \quad
      \psi
    = \pi / 2
\end{align}
with the positive sign in (\ref{eqn:remaining matching condition}).
In this case, $\lambda$ can easily be solved as
\begin{align}
    & \lambda
    = \sqrt{\frac{\pi}{2}\tan\left(\frac{\pi^2}{2\beta}\right)},
\end{align}
where $\beta = 2\pi / \mu_0$,
which gives the rotation symmetric 2-caloron of trivial holonomy obtained in \cite{Ward:2003sx}.
The constant surface of the action has a toroidal shape, as expected.
As the time coordinate $t$ varies, the surface still keeps its torus configuration:
When we increase $t$ from $t=0$, the action density gradually reduces its absolute value and 
almost fades away at 
$t=\beta/2$, and it returns to its initial value at $t=\beta$.

Now, we consider the general case accompanied with all the moduli parameters (\ref{eqn:moduli parameters}).
The quadruplets $d_1$, $d_2$, $d_3$, $d_0$ are the parameters of parallel transformation in $\mathbb{R}^3\times S^1$.
Substituting (\ref{eqn:2-caloron bulk data}) into (\ref{eqn:the Weyl equation}) 
one can easily see that for a parameter scaling $d_\mu\mapsto d_\mu+\delta_\mu$, $\delta_\mu$ 
is absorbed into $x_\mu$ as $x_\mu\mapsto x_\mu+\delta_\mu$.

It is already known for the BPS 2-monopoles that the modulus $k$ is a parameter of the geodesic motion of the monopoles.
The situation is similar for the 2-calorons.
In Fig. \ref{fig:isoaction_plot}, we present the action isosurface plots for several values of  $k = 0.0,0.7,0.9$,
which show the dynamical motion of the constituent monopoles.

Next, we consider $D$, which can be identified as a scale parameter of the caloron. 
Under a scale transformation with $\alpha$ being an arbitrary constant,
\begin{eqnarray}
D \mapsto \alpha D,~~s\mapsto s/\alpha,~~\lambda\mapsto \sqrt{\alpha}\lambda,~~\rho\mapsto \sqrt{\alpha}\rho,
\end{eqnarray}
the bulk  and the boundary Nahm data, (\ref{eqn:bulk solutions}),
(\ref{eqn:remaining matching condition}) are invariant.
If we fix the poles of the bulk data located on $s=\pm 1$, which gives a constraint on $D$ as $D=K(k)/2$,
then we can find a similar functional form for the bulk data as we change the value of $D$ and $\mu_0$, simultaneously.
This means that the change of $D$ leads the rescaling of the range of $I$, \textit{i.e.}, $\mu_0$. 
Now, the Weyl equation (\ref{eqn:the Weyl equation}) keeps invariant if the additional conditions are satisfied
\begin{eqnarray}
x_\mu\mapsto x'_\mu=\alpha x_\mu,~~U\mapsto U'(s/\alpha,x'_\mu)=\sqrt{\alpha}U(s,x_\mu)\,,
\end{eqnarray}
which obviously describe the spatial rescaling of the solution. 
Note that as $s$ reduces, $x_\mu$ increases at the same order. 
Hence, we conclude that $D$ works as a scale parameter of the solution.

The triplet, $\lambda$, $\rho$, and $\psi$, are the boundary data of the 2-caloron.
As discussed in \cite{Ward:2003sx}, {\it the symmetric calorons} are defined such that they are 
invariant under the group $G\subseteq \mathrm{SO}(3)$ as
\begin{align}
    \mathit\Theta_R^{- 1}(T_j \otimes \sigma^j)\mathit\Theta_R 
    &= T_j \otimes \sigma^j,~~\\
    \mathit\Theta_RW^\dagger
    &= W^\dagger\tau_R,
      \label{eqn:axially symmetric condition} %\ref{eqn:axially symmetric condition}
\end{align}
 where $\mathit\Theta_R = R_k \otimes R_2$ and $\tau_R$ is a unit quaternion.
 $R_2 \in \mathrm{SU}(2)$ is the rotation matrix acts on the quaternion coordinate. 
Similarly, $R_k$ is the $k$-dimensional irreducible representation of a rotation which acts on $k$-column vectors.
One can easily confirm that for rotations about $x_2$-axis, 
(\ref{eqn:axially symmetric condition}) is satisfied only for $\lambda = \rho,\psi=\pi/2$.
This means that the calorons are not always axially symmetric
 even if the bulk data is invariant.
We summarize figures of the action density to the non-axially symmetric 2-calorons
with the same bulk data to the rotation symmetric 2-calorons 
in Figs. \ref{fig:action_density_psi_rho1} and \ref{fig:action_density_psi_rho2}.

\begin{table} 
\centering
\caption{The role of each moduli parameter.}
\begin{tabular}{ccl}
    \hline\hline
        parameters                & ~ &\hspace{2.5cm}role \\
    \hline
        $d_1, d_2, d_3, d_0$      & ~ & Parallel transformation in $\mathbb{R}^3\times S^1$ \\
        $k$                       & ~ & Geodesic motion of two BPS monopoles \\
      $D$                       & ~ & Scale of the solution \\  
	$\phi$                    & ~ & Angle of the rotation around $x_2$-axis \\
        $\lambda$, $\rho$, $\psi$ & ~ & Parameters of the boundary data \\
    \hline\hline
\end{tabular}
\label{table:the roles of each parameters}%\ref{table:the roles of each parameters}
\end{table}

Finally, we consider  $\phi$, which describes a rotation around $x_2$-axis in $\mathbb{R}^3$, as mentioned in Section II.
This is because the bulk data can be written as (for positive sign)
\begin{eqnarray}
    & \left(\begin{array}{r} T_1(s) \\ T_2(s) \\ T_3(s) \end{array}\right)
    = \left(\begin{array}{ccc}
          \cos\phi & 0 &     - \sin\phi \\
                 0 & 1 & \quad        0 \\
          \sin\phi & 0 & \quad \cos\phi
      \end{array}\right)
      \left(\begin{array}{r}         
          a(s)\sigma_1   \\
          f_2(s)\sigma_2 \\
          b(s)\sigma_3
      \end{array}\right)
\end{eqnarray}
and $\phi$ can be set to $0$ by the rotation of quaternion $x_\mu e_\mu$ by $R_2$,
 together with the rotation on the Weyl spinors.
Thus, this parameter is dummy in the case of the 2-caloron of trivial holonomy. 
For the 2-caloron of non-trivial holonomy, however, this becomes a crucial parameter because 
it has a meaning of  the relative rotation angle of the constituent monopoles.

Hence, we have found full understandings for all parameters.
The results are summarized in Table \ref{table:the roles of each parameters}.

\section{Conclusion}
In this Letter, we have performed the Nahm transform numerically by directly solving the Weyl equations with boundary impurities
 for the 2-calorons of trivial holonomy, with 10 moduli parameters subject to 1 constraint.  
For the boundary equation, we have proposed the systematic method to extract appropriate 2-dimensional basis
 of the system by taking linear combinations of solutions of the Weyl equation.
The action density plots are qualitatively similar with those of \cite{Bruckmann:2004nu},
which are based on the Green function method.
The method presented in this Letter is essentially equivalent to \cite{Bruckmann:2004nu}. 

For the most general 2-calorons, with non-trivial holonomy, the program will also work well, 
though we need to solve more complicated matching conditions.
The role of the moduli parameters in this object is still not fully understood.
The analysis for such case is now in progress. 

The Atiyah-Manton construction of Skyrmions has already been generalized to
the case of finite temperature, by making use of calorons of 
trivial holonomy \cite{Eskola:1989qk,Nowak:1989gw}.  
For the study of a high energy collision of the nucleon or a high density phase of the nuclear matter, 
it will be valuable to consider the approximation to Skyrmions in more general situations. 
It is extremely interesting whether or not the calorons of non-trivial holonomy play a significant role 
in this subject. The analysis will be reported in forthcoming articles. 

\begin{acknowledgments}
The authors wish to thank the anonymous referee for careful reading 
of this manuscript and valuable remarks.
\end{acknowledgments}

\end{document}